\documentstyle[12pt]{article}
\begin{document}
\begin{titlepage}
\begin{flushright}
Z\"urich University ZU-TH 2/95\\
Pavia University FNT-T 95/2\\
Lecce University LE-ASTR 1/95
\end{flushright}
\vfill
\begin{center}
{\large\bf IS THE GALACTIC HALO BARYONIC ?}\\
\vfill
{\bf F. De Paolis$^{1,2,\#}$, G. Ingrosso$^{1,2,\#}$,
Ph.~Jetzer$^{3,}$* and M. Roncadelli$^4$}\\
\vskip 1.0cm
$^1$Dipartimento di Fisica, Universit\`a degli Studi di Lecce,
Via Arnesano CP 193,\\
 I-73100 Lecce, Italy,\\
$^2$INFN, Sezione di Lecce, Via Arnesano CP 193, I-73100 Lecce,
Italy,\\
$^3$Institute of Theoretical Physics, University of Z\"urich,
Winterthurerstrasse 190,\\
CH-8057 Z\"urich, Switzerland,\\
$^4$INFN, Sezione di Pavia, Via Bassi 6, I-27100 Pavia, Italy.
\end{center}
\vfill
\begin{center}
Abstract
\end{center}
\begin{quote}
Recent observations of microlensing events in the Large Magellanic
Cloud
suggest that a sizeable fraction of the galactic halo is in the form of
Massive Astrophysical Compact
Halo Objects (MACHOs) with mass less than about $0.1 M_{\odot}$.
Here we argue that molecular clouds (mainly of $H_2$)
located in the galactic halo
can contribute substantially to its total mass.
We outline a scenario in which dark clusters of MACHOs and molecular
clouds
naturally form in the halo at large
galactocentric distances.
Possible ways of detecting MACHOs via infrared
emission and molecular clouds via the induced $\gamma$-ray flux are
discussed.
Molecular clouds located in the M31 dark halo
could be discovered through cosmic background radiation (CBR)
anisotropies
or emission lines in the microwave band.

\noindent{
{\bf Key words:} dark matter, ISM: clouds, ISM: molecules, gamma rays:
theory,
Cosmic background radiation, Infrared radiation}
\end{quote}
\vfill
\begin{flushleft}
\vfill
\begin{center}
To appear in Comments on Astrophysics
\end{center}
\vskip 0.5cm
$^*$ Supported by the Swiss National Science Foundation.\\
$^{\#}$ Partially supported by Agenzia Spaziale Italiana.
\end{flushleft}
\end{titlepage}
\newpage

\baselineskip=21pt

One of the most important problems in astrophysics concerns the nature
of the
dark matter in galactic halos \cite{kn:Carr1},
whose presence is suggested by the observed  flat rotation
curves in spiral galaxies.
Although various
exotic dark matter candidates have been proposed,
present limits coming from primordial nucleosynthesis \cite{kn:Copi}
still allow a halo made of
ordinary baryonic matter. A viable candidate are
MACHOs in the mass range $10^{-7} < M/M_{\odot} < 10^{-1}$
\cite{kn:Derujula}, which can be detected, as proposed by Paczy\'nski,
via
the gravitational lens effect \cite{kn:Paczynski}. The few
microlensing events found so far \cite{kn:Alcock}
by monitoring stars in the Large
Magellanic Cloud (LMC)
do not yet allow to make a precise
estimate of the fraction of halo dark matter in the form of MACHOs nor
to infer
whether they are located in the halo or rather either in the LMC itself
\cite{kn:Sahu} or in a thick disk of our galaxy.
Assuming a standard spherical halo
model, Alcock et al. \cite{kn:Alcock1} found that MACHOs contribute
with a fraction $0.20^{+0.33}_{-0.14}$ to the halo dark matter, whereas
their
average mass turns out to be $\sim 0.08 M_{\odot}$ \cite{kn:Jetzer}.
Accordingly, the problem arises how to explain the
nature of the remaining fraction of the halo dark matter. {\it Here we
argue
that this fraction can still be baryonic and in the form of molecular
clouds (mainly of $H_2$)} \cite{kn:Depaolis}.
Actually, this point of view is corroborated by the result
\cite{kn:Moore}
that dissipationless particles, as advocated for cold dark matter,
can hardly make up the galactic halo. Moreover, it has recently
been claimed \cite{kn:Pfenniger} that $H_2$ molecular clouds can
constitute the dark matter in the disk of our galaxy.
Below, we present a scenario in which dark
clusters\footnote{The possibility of clusters of MACHOs
has been investigated by several authors (see e.g. \cite{kn:Carr}
-\cite{kn:Wasserman})}
of MACHOs and molecular clouds
naturally form at galactocentric distances $R$ larger than
10-20 kpc, basically because in a quite environment the Jeans
mass can drop to values as low as $10^{-2} M_{\odot}$.

Our picture encompasses the one originally proposed by Fall and Rees
\cite{kn:Fall} for the origin of stellar globular clusters and can be
summarized as follows.
After its initial collapse, the proto galaxy (PG) is expected to be
shock
heated to its virial temperature $\sim 10^6$ K. Since overdense regions
cool
more rapidly than average (by hydrogen recombination), proto globular
cluster
(PGC) clouds form in
pressure equilibrium with diffuse gas. At this stage, the PGC
cloud temperature is $\sim 10^4$ K while mass and size are
$\sim 10^6 (R/kpc)^{1/2} M_{\odot}$ and $\sim 10~(R/kpc)^{1/2}$ pc,
respectively.
Below $10^4$ K, the main coolants are $H_2$ molecules and any heavy
element
produced in a first chaotic galactic phase.
The subsequent evolution of the PGC
clouds will be different in the inner and outer part of the galaxy,
depending
on the decreasing collision rate and ultraviolet (UV) fluxes
as the galactocentric distance increases.

As discussed in \cite{kn:Fall,kn:Kang}, in the central region of the
galaxy
an Active Galactic Nucleus (AGN) and/or a first population of
massive stars are expected to exist,
which act as strong sources of UV radiation that dissociates
the $H_2$ molecules present in the inner part of the halo. As a
consequence,
cooling is heavily suppressed and so the PGC clouds remain for a long
time at temperature $\sim 10^4$ K, resulting in the imprinting
of a characteristic mass $\sim 10^6 M_{\odot}$.
Later on, when the UV flux decreases and after
enough $H_2$ forms, the cloud temperature suddenly drops below
$10^4$ K and the subsequent evolution leads to
the formation of stars and finally to stellar globular clusters.

Our main point is that in the outer regions of the halo the UV-flux
is suppressed due to the larger galactocentric distance,
so that no substantial $H_2$ depletion actually happens. On top of
this,
further $H_2$ is produced via three-body reactions
($H+H+H \rightarrow H_2+H$ and $H+H+H_2 \rightarrow   2H_2$), thus
the cooling efficiency increases dramatically.
This fact has three distinct implications:
(i) no imprinting of a characteristic PGC cloud mass shows up,
(ii) the Jeans mass can now be lower than $10^{-1} M_{\odot}$,
(iii)  the cooling time is much shorter than the collision time.
As pointed out in \cite{kn:Palla}, a subsequent fragmentation
occurs into smaller clouds that remain optically thin until the minimum
value
of the Jeans mass is attained, thus
leading to MACHO formation. Moreover, since the
conversion efficiency of the constituent gas could scarcely have been
100\%,
we expect the remaining fraction $f$ of the gas to form
self-gravitating
molecular clouds since, in the absence of
strong stellar winds, the surviving gas
remains bound in the dark cluster. The possibility that the
gas is in diffuse
form either in the dark cluster or in the galactic halo
is in fact excluded, as the high
virial temperature would make it observable in radio or X-ray band.

A few comments are in order. Because the formation of dark
clusters of MACHOs and molecular clouds requires sufficiently low UV
fluxes,
they can mainly form beyond a critical galactocentric
distance, which we estimate
to be $R_{crit} \sim 10-20$ kpc \cite{kn:Depaolis}.
Obviously, the
above discussion implicitly assumes that dark clusters are stable
within the
lifetime of the galaxy. This is a nontrivial question, for dark
clusters can
be disrupted by evaporation and collisions among themselves, whereas
molecules can be destroyed by strong UV fluxes.
All these effects are avoided, provided the
galactocentric distance of dark clusters exceeds $R_{dis} \sim 10$ kpc
and
both MACHO and molecular cloud masses are less than about $0.1
M_{\odot}$
\cite{kn:Carr}.

Let us now briefly discuss the possible signatures of the above
scenario.
The most
promising way to detect dark clusters of MACHOs is via correlation
effects
in microlensing observations, as they are expected to exhibit a
cluster-like
distribution. Remarkably enough, a relatively small number of
microlensing
events
would be sufficient to rule out this possibility, while to confirm it
more
events are needed \cite{kn:Maoz}.

A signature of the presence of molecular clouds in the galactic halo
should be a $\gamma$-ray flux produced through the interaction with
high-energy
cosmic ray protons which, scattering on $H_2$ protons, produce
$\pi^0$'s which
subsequently decay into $\gamma$'s. As a matter of fact, an essential
ingredient is the knowledge of the cosmic ray flux in the halo.
Unfortunately,
this quantity is unknown and the only available
information comes from theoretical estimates.
More precisely,
from the mass-loss rate of a typical galaxy, we infer a total cosmic
ray flux
in the halo $F \simeq 1.1\times 10^{-4}$ erg cm$^{-2}$ s$^{-1}$. We
further
assume the same energy distribution of the cosmic rays as measured on
Earth and
we scale the cosmic ray density with the inverse of $R^2$.
Actually, cosmic ray protons in the halo originating from
the galactic disk are mainly directed outwards. This circumstance
implies that
the induced photons will predominantly leave the galaxy.
However, the presence of
magnetic fields in the halo could give rise to a temporary confinement
of
cosmic ray protons similar to what happens in the disk. In addition,
there
can also be sources of cosmic ray protons located in the halo itself,
as for
instance isolated or binary pulsars in stellar globular clusters.
Since we are unable to give a quantitative estimate of the above
effects, we take them into account by introducing an
efficiency factor $\epsilon$, which could be rather small.
The best chance to detect the $\gamma$-rays in question is provided
by observations at high galactic latitude.
Accordingly, we find a $\gamma$-ray flux
$\Phi_{\gamma}(90^0) \simeq ~3 \times 10^{-6}~\epsilon f$ photons
cm$^{-2}$
s$^{-1}$  sr$^{-1}$  \cite{kn:Depaolis,kn:Silk},
while the inferred upper bound for
$\gamma$-rays in the 0.8 - 6 GeV range at high galactic latitude is
$3 \times 10^{-7}$ photons cm$^{-2}$ s$^{-1}$  sr$^{-1}$. Thus,
the presence of molecular clouds in the galactic halo does not lead at
present
to any contradiction with the upper bound, provided $\epsilon f <
0.1$.

A perhaps better way to discover the molecular clouds in question
relies
upon their emission in the microwave band.
The temperature of the clouds is close to that of the cosmic background
radiation (CBR). Indeed, an upper limit of $\Delta T/T \sim 10^{-3}$
can
be derived \cite{kn:Depaolis3} by considering the anisotropy they would
introduce in the CBR due to their higher temperature. Realistically,
molecular
clouds cannot be regarded as black body emitters because they mainly
produce a set
of molecular rotational transition lines.
If we consider
clouds with cosmological primordial composition, the only molecule that
contributes to the microwave band with optically thick lines is LiH
\cite{kn:Melchiorri}, whose lowest rotational
transition occurs at $\nu_0 = 444$ GHz with broadening $\sim 10^{-5}$
(due to the turbulent velocity of molecular clouds in dark clusters).
This line would be detectable using the Doppler shift effect.
To this aim, it is convenient to consider the M31 galaxy, for
whose halo we assume the same picture as outlined above for our galaxy
\cite{kn:note}.
Then we expect that molecular clouds should
have typical rotational speeds of 50-100 km s$^{-1}$.
Given the fact that the clouds possess a peculiar velocity
with respect to the CBR, the
emitted radiation will be Doppler shifted with
$\Delta\nu /\nu_{0}\sim\pm 10^{-3}$.
However, the precise chemical composition of molecular clouds in the
galactic halo is unknown. Even if the heavy
molecule abundance is very low (as compared with the abundance in
interstellar clouds), many optically thick lines corresponding to the
lowest
rotational transitions would show up in the microwave band. In this
case,
it is more convenient to perform broad-band measurements, because
molecular
clouds may be discovered using again the Doppler shift effect, thereby
producing an anisotropy in the CBR itself.
Since it is difficult to work with fields of view
of a few arcsec, we propose to measure the
CBR anisotropy between two fields of view - on opposite
sides of M31 - separated by $\sim 4^0$ and with angular resolution
of $\sim 1^0$. We suppose that the halo of M31 consists of
$\sim 10^6$ dark clusters which lie within 25-35 kpc.
Scanning an annulus of $1^0$ width and internal angular
diameter $4^0$, centered at M31, in 180 steps of $1^0$, we would find
anisotropies of $\sim 10^{-5} ~f~ \bar\tau$ in $\Delta T/T$
\cite{kn:Depaolis3}.
Here, most of the
uncertainties arise from the estimate of the
average optical depth $\bar\tau$, which mainly depends
on the molecular cloud composition. In conclusion, since the theory
does not
allow to establish whether the expected anisotropy lies above or below
current detectability ($\sim 10^{-6}$), only observations can resolve
this
issue.

For completeness, we mention that another possibility of detecting
MACHOs
is via their infrared
emission \cite{kn:Adams}.
In order to be specific, let us assume that all
MACHOs have same mass 0.08 $M_{\odot}$ and age
$10^{10}$ yr. Accordingly, their surface temperature is
$\sim 1.4 \times 10^3$ K and they emit most of their radiation
(as a black body) at $\nu_{max} \sim 11.5 \times 10^{13}$ Hz.
First, we consider MACHOs  located in M31.
In this case, we find a surface brightness
$I_{\nu_{max}} \sim 1.6 \times 10^3~(1-f) $ Jy sr$^{-1}$ and
$0.4 \times 10^{3}~(1-f)$ Jy sr$^{-1}$
for projected separations from the M31 center $b=20$ kpc and 40 kpc,
respectively \cite{kn:Depaolis3}.
Although these values are about one order of magnitude below the
sensitivity of
the detectors on ISO Satellite, they lie above the threshold of
the future planned satellite SIRFT.
For comparison, we recall that the halo of our galaxy would have in
the direction of the galactic pole a surface brightness
$I_{\nu_{max}}\sim 2 \times 10^{3}~{\rm Jy~sr^{-1}}$, provided
MACHOs make up the total halo dark matter.
Nevertheless, the infrared radiation
originating from  MACHOs in the halo of our galaxy can be recognized
(and
subtracted) by its characteristic angular modulation.
Also, the signal from the
M31 halo can be identified and separated from the galactic background
via its
b-modulation.
Next, we point out that the angular size of dark clusters in the
halo of our galaxy at a distance of $\sim 20$ kpc is $\sim 1.8'$ and
the
typical separation among them is
$\sim 14'$. As a result, a characteristic pattern of
bright (with intensity $\sim 3\times 10^{-2}$ Jy at $\nu_{max}$ within
angular size $1.8'$  \cite{kn:nota1}) and dark spots
should be seen by pointing the detector into
different directions.

We would like to close this paper by suggesting out that the above
scenario might also hold for elliptical galaxies, for which there is
evidence
for the presence of dark matter (see e.g. \cite{kn:Depaolis4}).
Also in this case, the dark matter can be in the form of dark clusters
of MACHOs and molecular clouds, which are not destroyed by the
X-ray flux (whose intensity is always less than $\sim 10^{43}$ erg
s$^{-1}$)
present in ellipticals.
An advantage of this idea is that the diffuse gas observed in clusters
of galaxies can be understood as arising from the tidal stripping from
neighboring galaxies.
This fact naturally explains the ROSAT observation that
$\sim 30$\% of the dynamical mass in clusters of galaxies is in
baryonic form
\cite{kn:Chi}.
These speculations would lead to a unique picture for galaxy
formation.
The parameter which mainly discriminates between the formation of
either
spirals or ellipticals would be the initial total angular momentum.


\begin{thebibliography}{99}
\bibitem{kn:note}
Here we suppose, for
illustration, that dark clusters are spherically distributed at $\sim
20$ kpc
from the galactic center and consist of $10^6 (M_{\odot}/m) f$
molecular clouds
(each of mass $m$) and radius (from the virial theorem)
$\sim 2 \times 10^{-1} (m /M_{\odot})$ pc.
\bibitem{kn:nota1}
As an orientation, we mention that this intensity is within
the sensitivity of the
detector ISOPHT aboard on ISO satellite.
\end{thebibliography}
\end{document}